\documentclass[namedreferences]{solarphysics}
\usepackage[hyperref,optionalrh,solaromanenum]{spr-sola-addons} 
\usepackage[optionalrh,solaromanenum]{spr-sola-addons}

\usepackage{graphicx}                   
\usepackage{breakurl}                         
\usepackage{comment}

\begin{document}

\begin{article}

\begin{opening}

\title{Development of active regions: flows, magnetic-field patterns and bordering effect\\
{\it Solar Physics}}

%
\author[addressref={avg},corref,email={A.Getling@mail.ru}]{\inits{A.V. }\fnm{A. V. }\lnm{Getling}}
\author[addressref={ri},email={ryoko.ishikawa@nao.ac.jp}]{\inits{R. }\fnm{R. }\lnm{Ishikawa}}
\author[addressref={aab},email={baa@ooi.sscc.ru}]{\inits{A.A. }\fnm{A.~A.~}\lnm{Buchnev}}

%
\runningauthor{A. V. Getling, \emph{et al.}} \runningtitle{Development of
active regions}

\address[id={avg}]{Skobeltsyn Institute of Nuclear Physics, Lomonosov Moscow State
University, Moscow, 119991 Russia}
\address[id={ri}]{Hinode Science Center, National Astronomical Observatory of
Japan, 2-21-1 Osawa, Mitaka, Tokyo, 181-8588 Japan}
\address[id={aab}]{Institute of Computational Mathematics and Mathematical Geophysics,
Novosibirsk, 630090 Russia}

\begin{abstract}
A qualitative analysis is given to the data on the full magnetic and velocity
vector fields in a growing sunspot group, recorded nearly simultaneously with
the \textit{Solar Optical Telescope} on the \emph{Hinode} satellite.
Observations of a young bipolar subregion developing within AR 11313 were
carried out on 9-10 October 2011. Our aim was to form an idea about the
consistency of the observed pattern with the well-known rising-tube model of
the formation of bipolar active regions and sunspot groups. We find from our
magnetograms that the distributions of the vertical [$B_\mathrm v$] and the
horizontal [$B_\mathrm h$] component of the magnetic field over the area of the
magnetic subregion are spatially well correlated; in contrast, the rise of a
flux-tube loop would result in a qualitatively different pattern, with the
maxima of the two magnetic-field components spatially separated: the vertical
field would be the strongest where either spot emerges, while the maximum
horizontal-field strengths would be reached in between them. A specific
feature, which we call the \emph{bordering effect}, is revealed: some local
extrema of $B_\mathrm v$ are bordered with areas of locally enhanced $B_\mathrm
h$. This effect suggests a fountainlike spatial structure of the magnetic field
near the $B_\mathrm v$ extrema, which is also hardly compatible with the
emergence of a flux-tube loop. The vertical-velocity field in the area of the
developing active subregion does not exhibit any upflow on the scale of the
whole subregion, which should be related to the rising-tube process. Thus, our
observational data can hardly be interpreted in the framework of the
rising-tube model.
\end{abstract}

%

\end{opening}

\section{Introduction}

In the developing active regions (ARs), subphotospheric magnetic fields and
fluid motions are strongly coupled according to the laws of
magnetohydrodynamics. Obviously, the new magnetic flux emerges to the solar
surface from deeper layers; if the emerging field is sufficiently strong,
sunspot groups may develop. However, an intriguing issue related to this
process is whether the strong magnetic field plays an active, primary role,
having been prepared in the depth of the convection zone and then emerging
through the visible surface of the Sun, or the primary role is played by the
fluid motion, which amplifies and structures the initially moderate or weak
magnetic field.

The first of these alternatives is typically represented by the popular
\emph{rising-tube model} (RTM) according to which the magnetic field of a
bipolar sunspot group originates from the emergence of an $\Omega$-shaped loop
of a coherent flux tube of strong magnetic field. As usually assumed, the flux
tube forms in the general toroidal\footnote{As frequently done in the
literature on stellar and planetary dynamos, we use here the terms
\emph{toroidal} and \emph{azimuthal} as synonyms, although they are not
mathematically equivalent.} solar magnetic field deep in the convection zone,
and the field that this tube carries upward is already strong before the rise.
This mechanism received much attention after a well-known study by
\citet{parker}, who invoked magnetic buoyancy to account for the rise of
flux-tube loops. Later, the RTM has been revisited by a number of investigators
over several decades. Interesting considerations of this model were suggested,
in particular, by \citet{clgretal1} and \citet{clgretal2}; numerical
simulations of this mechanism based on full systems of MHD equations have also
been carried out \citep[see, e.g.,][and references
therein]{fanetal,rempcheung}.

The concept of RTM, which agrees with such important regularities of solar
activity as Hale's polarity law and Sp\"orer's law of sunspot-formation
latitudes, appeared to be very attractive in the epoch of moderate capacities
of observational instrumentation. For this reason, the idea of RTM remained
virtually indisputable for a long time. Currently, the RTM still receives
attention in both analyses of observational data and numerical simulations. In
particular, under the assumption that a twisted flux tube (rope) rises in a
developing AR, \citet{Luoni_etal:2011} and \citet{Poisson_etal:2015} estimate
the magnetic helicity using the configurations of magnetic polarities in
observed ARs.

However, some implications of the RTM were found to be hardly compatible with
observations. In particular, if a flux-tube loop is rising, two very impressive
manifestations of this process should definitely be observed but, as we shall
see, they do not conform with the observations discussed below.
\begin{enumerate}
  \item As the loop is rising, strong horizontal magnetic fields should
      emerge on the scale of the entire AR. It is the pattern of magnetic
      fields in a developing AR that will be the subject of our discussion
      here, and we shall find no manifestations of the loop rise.
  \item Intense spreading from the site of the loop emergence should be
      observed over the whole AR. This is actually not the case; in
      particular, \citet{PevtsovLamb:2006} ``observed no consistent plasma
      flows at the future location of an active region before its
      emergence'' and \citet{kosov} finds ``no evidence for large-scale
      flows indicating future appearance a large-scale magnetic structure
      without signs of a large-scale horizontal flow''. A further example
      of the horizontal-velocity pattern without spreading on the scale of
      the whole developing AR is given by \citet{getling_etal_hinode}
      (hereinafter, Paper~I).
\newcounter{lenumi}\setcounter{lenumi}{\value{enumi}}
\end{enumerate}

\noindent In addition, it is worth mentioning two other important
considerations that are also not in favour of the RTM:

\begin{enumerate}\addtocounter{enumi}{\value{lenumi}}
  \item If the RTM is adopted, one has to account for the origin of the
      coherent tube of strong magnetic field deep in the convection zone.
      Various assumptions have been made to this end, which differ in their
      plausibility and the appropriateness of their starting points
      \citep[see, e.g., a review by][and references therein]{Fan:2009-4}.
      Furthermore, before the emergence on the photospheric surface, such
      an intense flux tube should affect the structure of the convective
      velocity field, which also is not actually observed.
  \item Joy's law of the latitudinal dependence of the tilt angle of
      bipolar sunspot groups is not sensitive to the amount of the emerging
      magnetic flux in contrast to what could be expected: as
      \citet{kosovstenflo} and \citet{kosov} write, their ``new statistical
      study of the variations of the tilt angle of bipolar magnetic regions
      during the flux emergence \emph{questions the current paradigm} that
      the magnetic flux emerging on the solar surface represents
      large-scale magnetic flux ropes ($\Omega$-loops) rising from the
      bottom of the convection zone'' (italicised by us. -- Authors).
\end{enumerate}

These are the most important points of doubt about the universal applicability
of the RTM; we shall not go here into further details, since we already
discussed some of them in Paper~I. In view of the contradictions of the RTM
with observations and difficulties of accounting for the origin of the intense
flux tube and some other features of the process, the rising-tube mechanism no
longer appears to determine a paradigm in the studies of the development of
ARs.

As alternatives to the RTM, various mechanisms of \emph{in situ} magnetic-field
amplification and structuring have been suggested. Among them, MHD mechanisms
of inductive excitation of magnetic fields strongly coupled with fluid motions
are usually referred to as \emph{local dynamos}. This term is frequently used
only in the context of small, granular scales and under; in contrast, we
associate the concept of local dynamo with a wider range of scales, including
the sizes of the whole ARs (mesoscales).

The idea of local MHD dynamo traces back to \citet{gurleb}, who related the
amplification process to the effects of plasma motions; however, they did not
attribute these motions to convection and even did not specify any particular
type of motion. \citet{tver} used a very simple model to demonstrate that the
magnetic field can locally be amplified and structured by cellular
magnetoconvection. Later, numerical simulations revealed the role of local
dynamos as the producers of fairly disordered, intermittent magnetic fields on
very small (down from the granular) scales \citep[][
etc.]{cattaneo,voeglschuessl,kitkos:2015} In essence, only \citet{steinnord}
used ``realistic'' numerical simulations to describe a convective mechanism
capable of producing mesoscale amplified magnetic fields. They investigated the
formation of an AR via the flux rise due to convective motions in the upper
portion of the convection zone. The computed scenario does not imply the
pre-existence of a coherent flux tube. A uniform, untwisted, horizontal
magnetic field is initially present, and magnetic loops subsequently form over
a wide range of scales. However, the initial field is required to be relatively
strong, and only a moderate magnetic-field amplification (by a factor of about
three) can be achieved. Previously, we discussed local convective dynamos in
more detail (see Paper~I).

Sunspot-formation mechanisms differing from local dynamo were also suggested.
\citet{kitmaz} investigated a hydromagnetic instability that can act on scales
large compared to the granular size, producing a magnetic-flux concentration
similar to those observed in sunspots. The process crucially depends on the
presence of fluid motion and on the quenching of eddy diffusivity by the
enhanced magnetic field with the plasma cooling down. This is a local
mechanism, which, however, is not a dynamo in the strict meaning of this term.

Another local mechanism, which does not qualify as a dynamo, is related to
the so-called negative-effective-magnetic-pressure instability (NEMPI), see
\citet{warnetal2013,warnetal2015} and references therein. It results from the
suppression of the total turbulent pressure (the sum of hydrodynamic and
magnetic components) by the magnetic field.

We present here a qualitative consideration of some observational data from the
standpoint of items (i) and (ii) in the above list to decide whether or not
these data can be interpreted in terms of the RTM. To obtain relevant data, we
developed an observational program of studying the evolution of both the
velocity and magnetic fields in growing ARs. This program (operation plan) is
intended for implementation with the \textit{Solar Optical Telescope} (SOT) on
the \emph{Hinode} spacecraft \citep{tsunetaetal,suematsuetal,shimizuetal} and
has been designated as HOP181
(\url{http://www.isas.jaxa.jp/home/solar/hinode\_op/hop.php?hop=0181}). It
consists in simultaneously recording and analysing the dynamics of the
full-vector velocity and magnetic fields on the photospheric level. Previously,
we presented some preliminary results of our study (Paper~I). New features of
the AR evolutionary pattern inferred from our data will be given here.

\section{Observations and Data Processing}\label{obs}

A bipolar magnetic structure, which emerged within AR~11313, was observed at
its early evolutionary stage, on 9--10 October 2011; the AR was then near the
centre of the solar disc. Five 2-h-long observational sessions were carried out
with intervals that varied from 3 h 40 min to 6 h 30 min (see a summary in
Table~\ref{summary}).

\begin{table}[h]
\caption{Summary of observational sessions}\label{summary}
\begin{tabular}{cccc}
  \hline
  Session & Date & Session & Interval \\
  No.    &      & meantime & from (1) \\
  \hline
  1 & 09 Oct. 2011 & 19:31:15 & 00:00
 \\
  2 & 10 Oct. 2011 & 01:06:16 & 05:35
 \\
  3 &              & 06:30:15 & 10:59
 \\
  4 &              & 15:00:14 & 19:29 \\
  5 &              & 21:46:15 & 26:15
 \\
  \hline
\end{tabular}
\end{table}

During each session, a $150''\times 163''$ field of view (FOV) was observed
using the \textit{Narrowband Filter Imager} (NFI) of the SOT at two wavelength
positions of Fe{\sc I} $\lambda$~5776~\AA\ with a time cadence of 2~min and a
pixel size of $0.16''$. This yielded a series of photospheric images, which can
be used to calculate horizontal\footnote{Since the area of interest was located
near the solar-disc centre and, moreover, corrections for projection effects
are not important from the standpoint of our goal, we do not make difference
here between the line-of-sight and the vertical component and also between the
transversal (tangential) and horizontal vector components.} velocities
[$u_\mathrm h$], and a series of Dopplergrams representing the line-of-sight
(LOS), or vertical, velocities [$u_\mathrm v$]. Simultaneously, the same FOV
was scanned with the \textit{Spectro-Polarimeter} \citep[SP;
see][]{ichimotoetal,litesetal13} one or two times a session. The SP scan was
done in the so-called fast mode with a pixel size of $0.32''$, taking 32 min to
obtain one SP map. To derive full-vector magnetic fields from these SP
observations, we used the MERLIN inversion code \citep{litesetal07}, which
assumed a Milne--Eddington atmosphere.

The processing of the data included:
\begin{enumerate}
\item Subsonic filtering based on Fast Fourier Transform;
\item Constructing Dopplergrams;
\item An intensity-scaling procedure enhancing the image contrast by means
    of cutting off the tails of a pixel-intensity histogram and subsequent
    linear mapping of the remaining portion of the histogram onto the whole
    admissible intensity range;
\item Alignment of the magnetograms corrected for irregularities of the SP
    scanning process, with the properly rescaled images and Dopplergrams
    obtained at the meantime of the scan (the correction was done by
    choosing 30 or more reference points, typically pores or easily
    identifiable fine details of spots, in both the FG image and the image
    accompanying the magnetogram, and subsequently bringing them into
    coincidence by means of affine transformations);
\item Determination of the horizontal-velocity field using a technique
    based on the same principle as the standard method of local correlation
    tracking (LCT) but more reliable \citep[see][for a description]{gbuch}
    and construction of cork-trajectory maps. Our technique differs from
    the standard LCT procedure in a special choice of trial areas
    (``targets''), whose displacements are determined by maximizing the
    correlation between the original and various shifted positions of the
    target. Specifically, an area is chosen as a target in a certain
    neighborhood of each node of a predefined grid if either the contrast
    or the entropy of the brightness distribution reaches its maximum in
    this area. The horizontal velocities obtained are then interpolated to
    the positions of imaginary ``corks'' using the Delaunay triangulation
    and affine transformations specified by the deformation of the obtained
    triangles at the time step considered;
\item Elimination of the Sun's rotation from the fields of LOS velocities;
\item Reducing the mean LOS velocity to zero in each map.
\end{enumerate}

The MERLIN code yields the magnitude, inclination and azimuth of the
magnetic-field vector (in parallel with a continuum image), which we then
convert into the LOS and tangential components. In addition to the original
maps of the vertical velocity and magnetic-field components, we used smoothed
maps of these quantities.

\begin{figure} 
\centering
\includegraphics[width=0.495\textwidth]{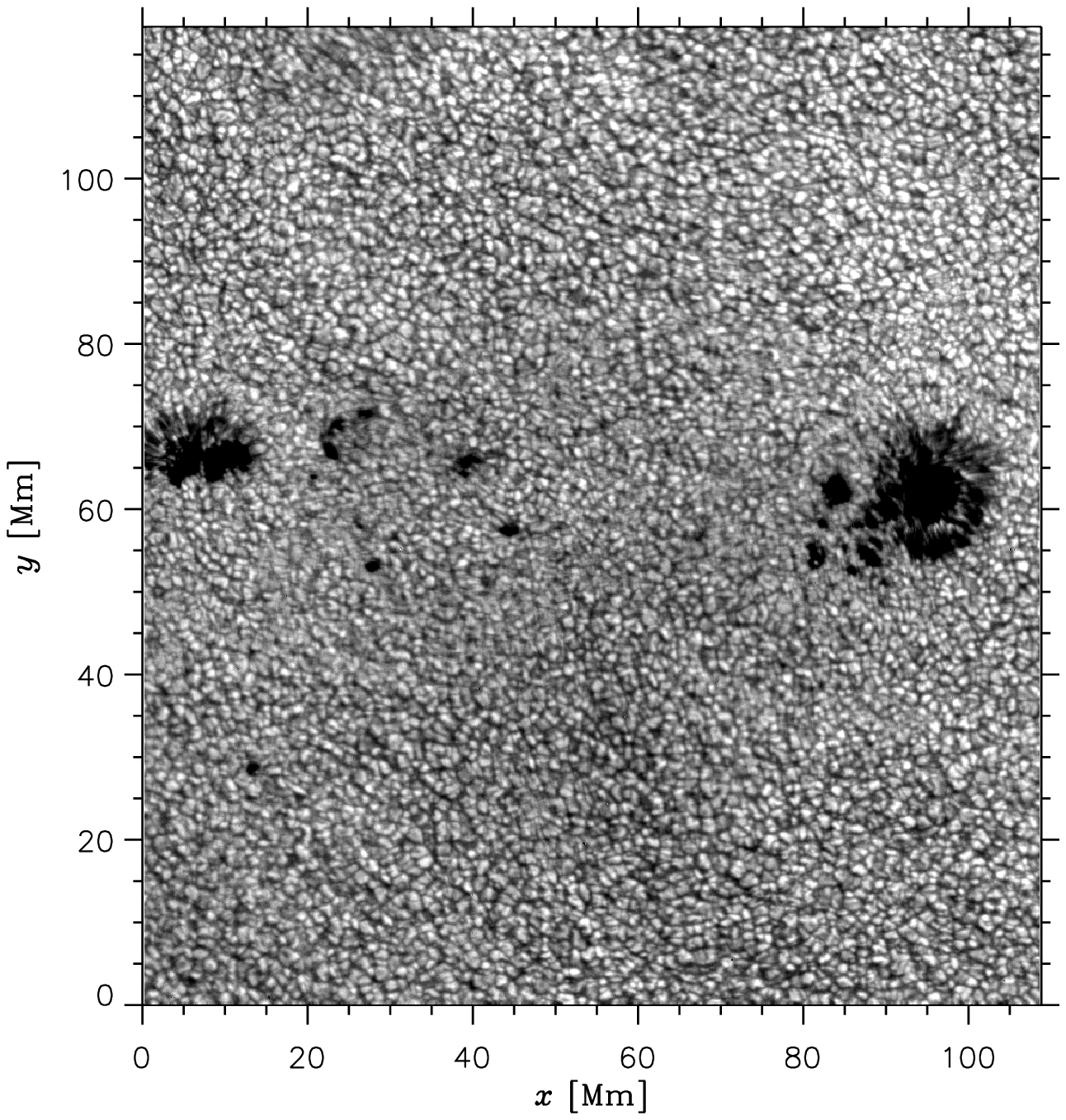}
\includegraphics[width=0.495\textwidth]{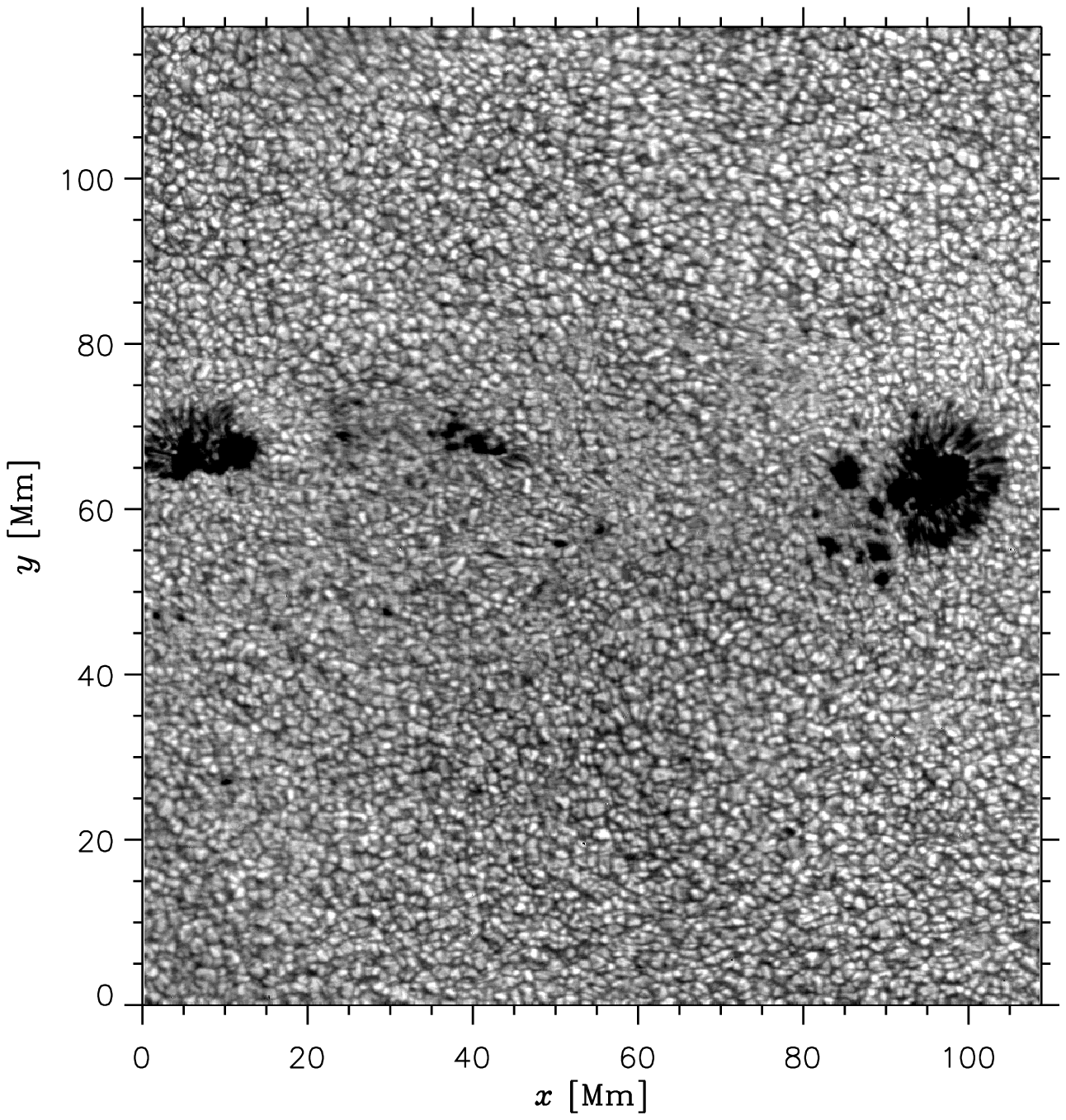}\\
\small{\hspace{0.7cm}(a) \hspace{5.5cm} (b)}\\
\includegraphics[width=0.495\textwidth]{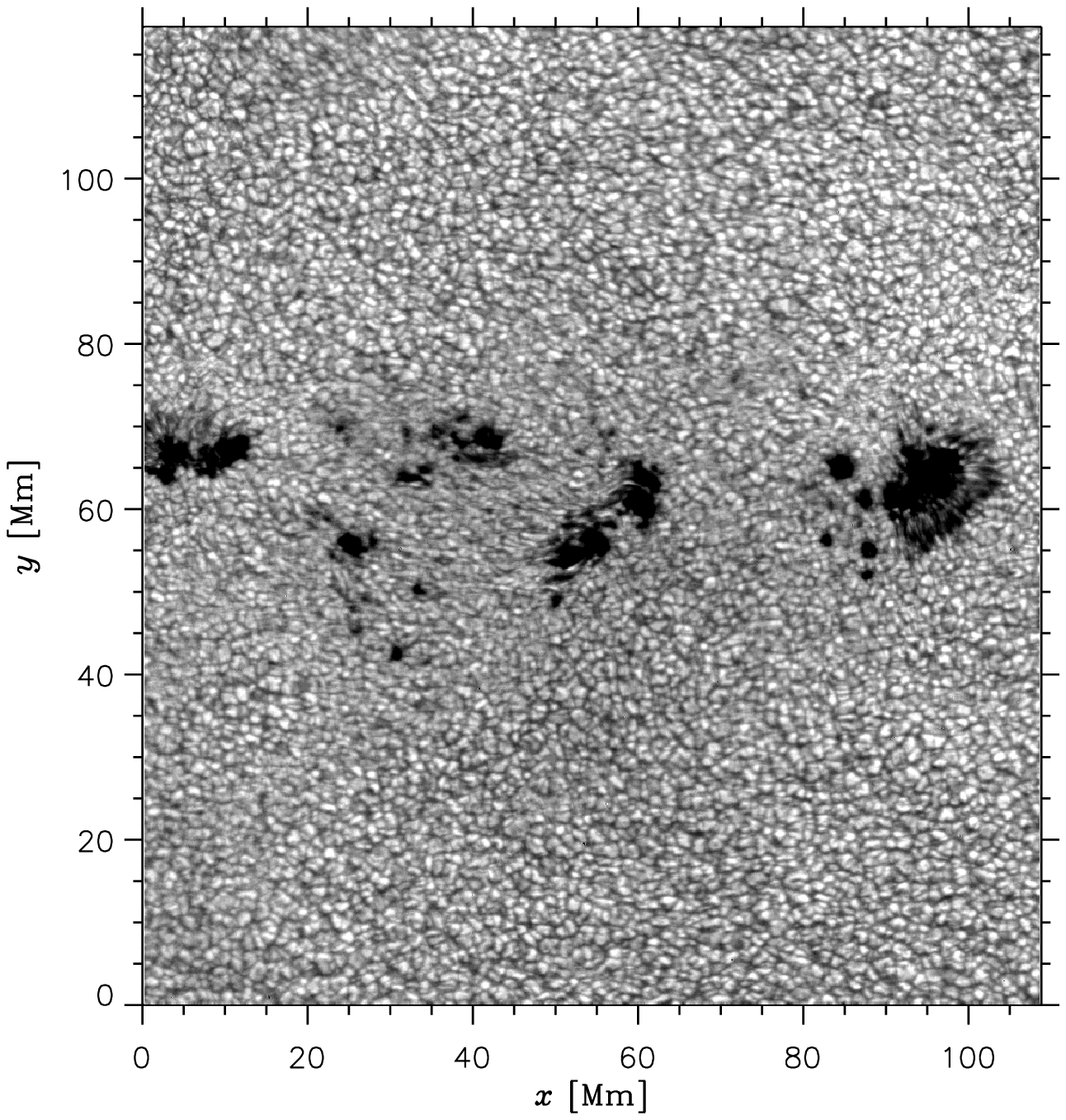}
\includegraphics[width=0.495\textwidth]{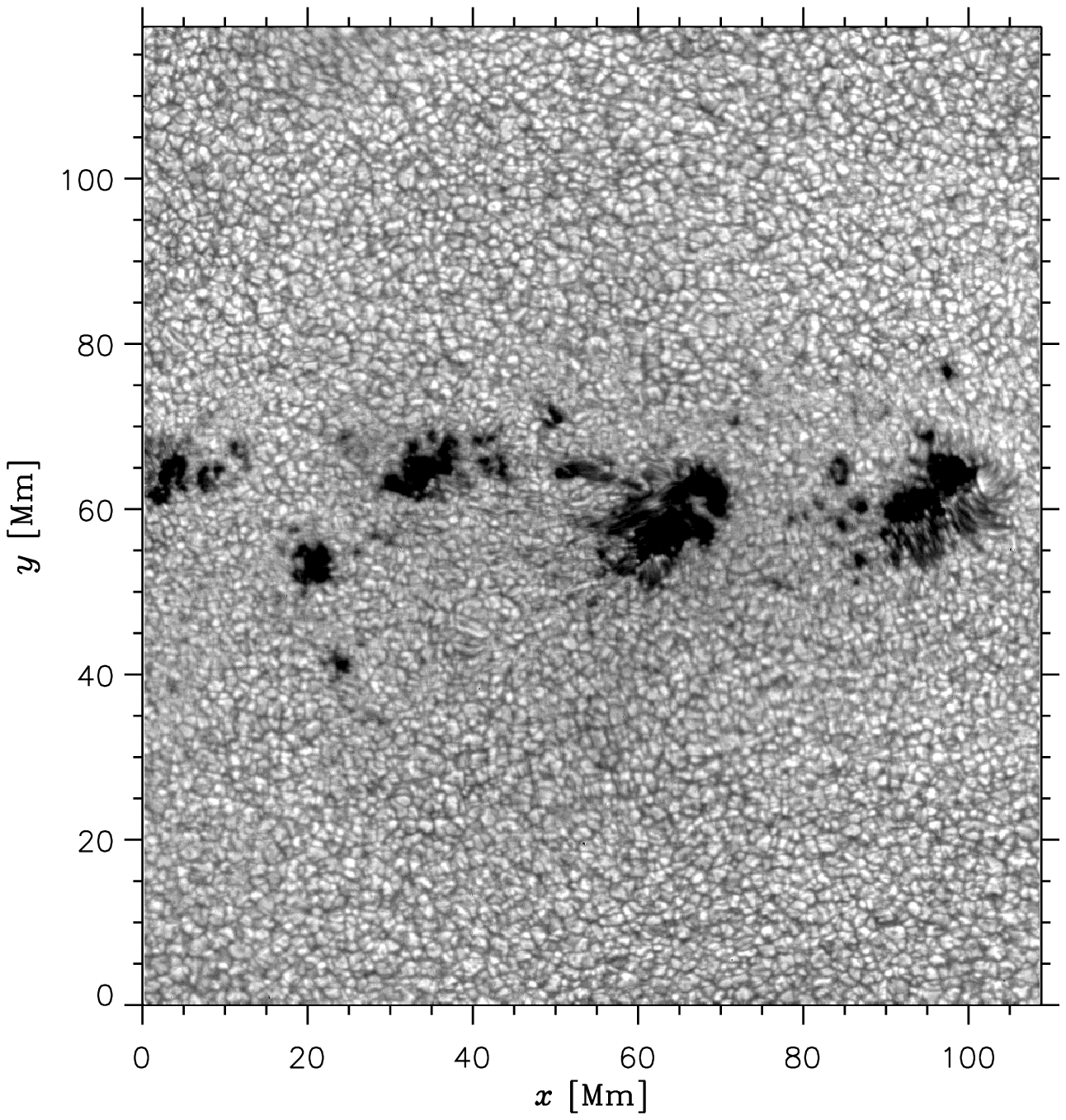}\\
\small{\hspace{0.7cm}(c) \hspace{5.5cm} (d)}\\
\includegraphics[width=0.495\textwidth]{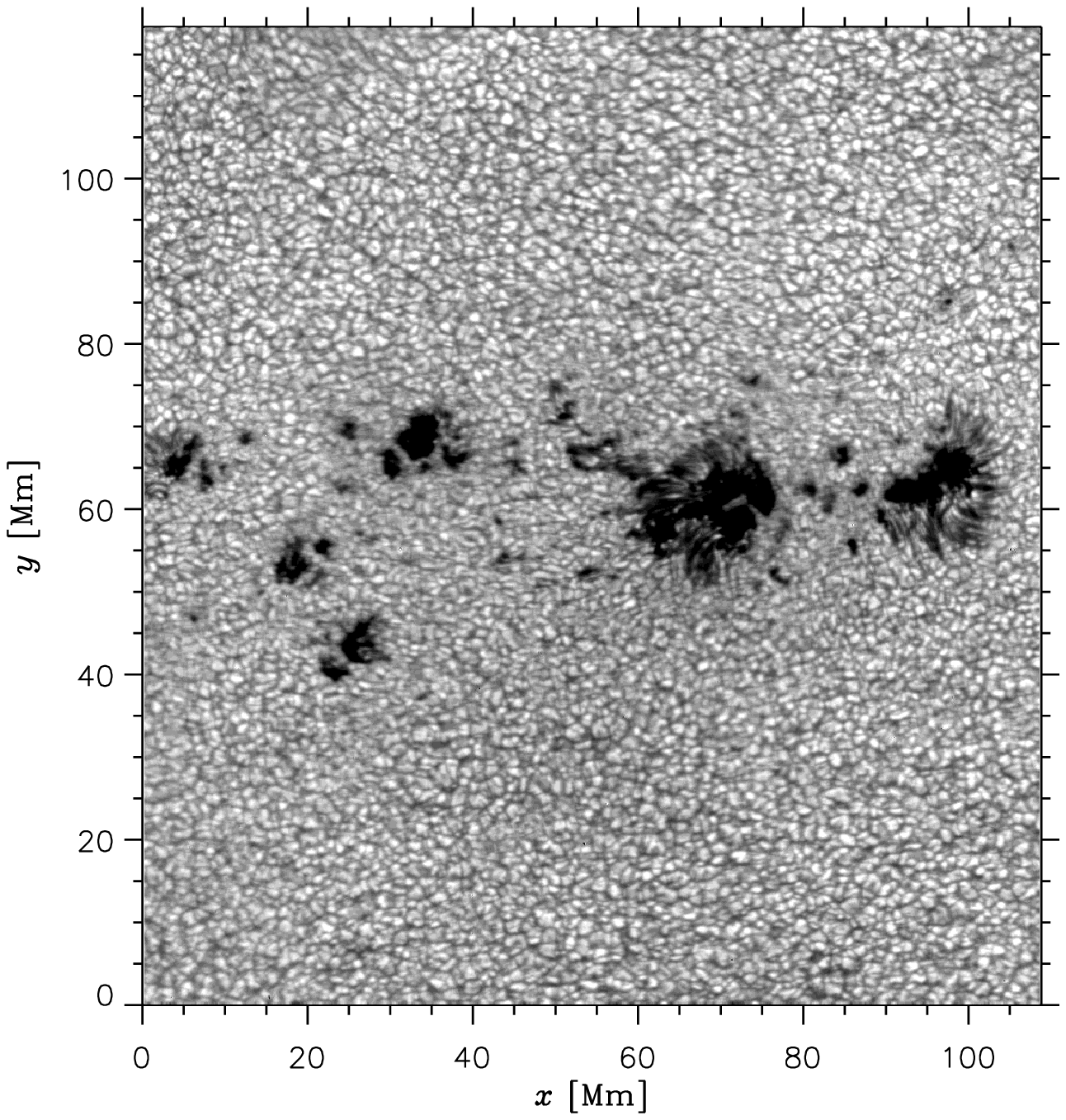}\\
\small{\hspace{0.7cm}(e)}
\caption{Evolution of AR 11313 in visible light. Images (a)--(e) were taken at the mean times
of observational sessions (1)--(5) (see Table~\ref{summary}).\protect} \label{images}
\end{figure}

\begin{figure}[t!]
\centering
\includegraphics[width=0.495\textwidth,bb=0 0 377 411,clip]{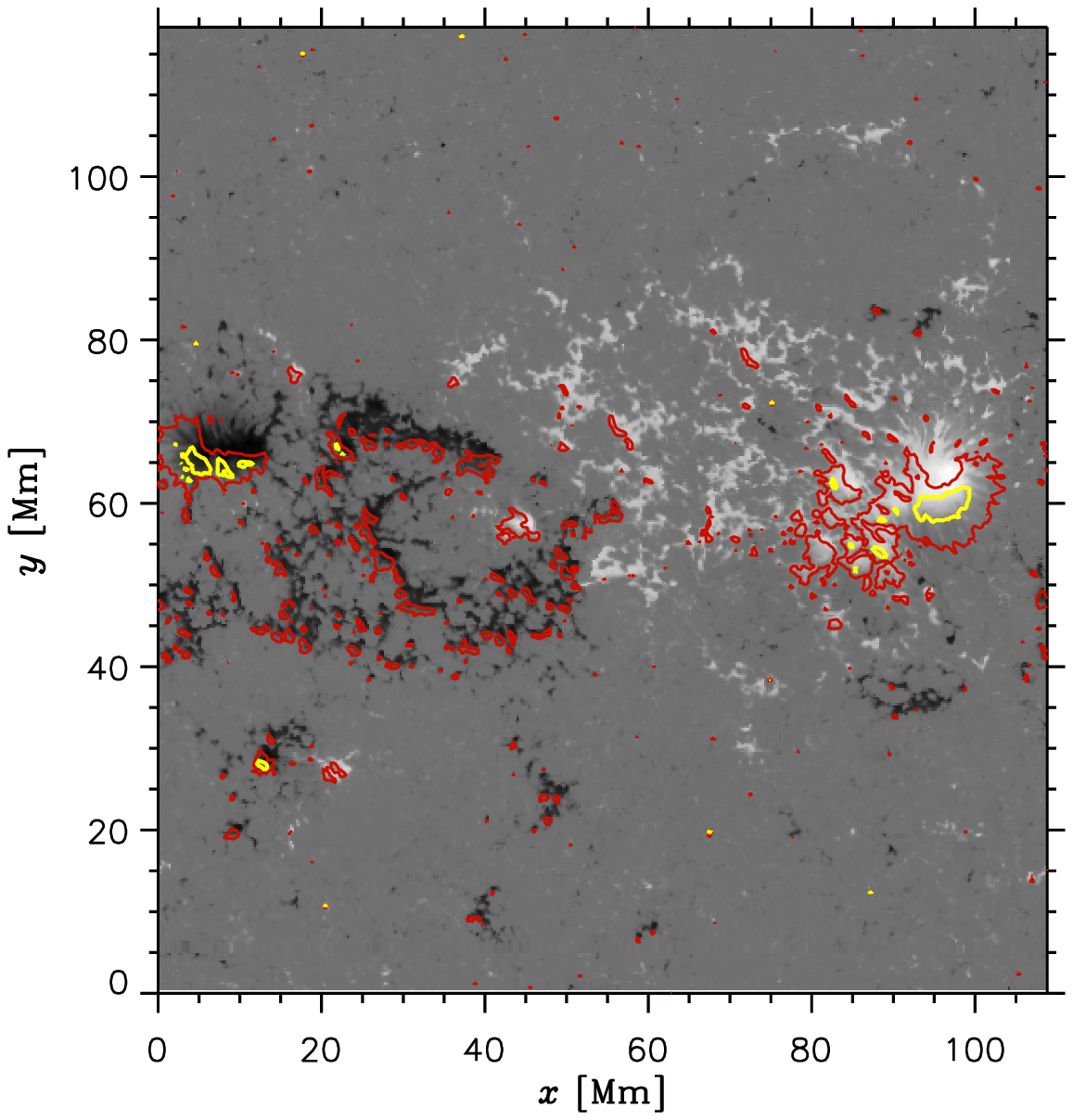}
\includegraphics[width=0.452\textwidth,bb=15 0 360 360,clip]{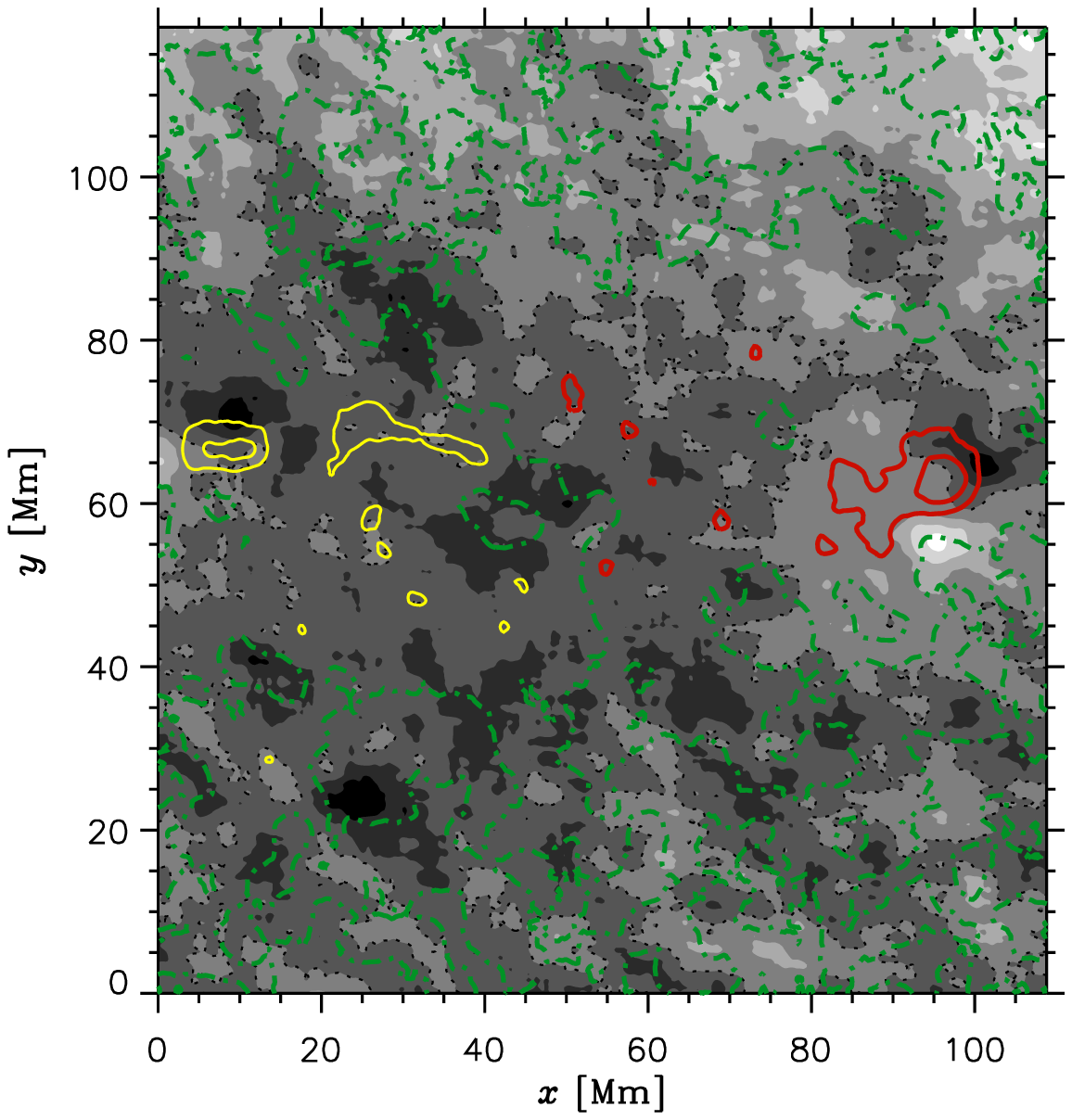}
\caption{Left: comparison between the vertical (grey-scale map) and horizontal (contours) components
of the magnetic field for observational session (1); the $B_\mathrm v$ range is $[-1941~\mathrm G,\
2477~\mathrm G]$;
contour levels for $B_\mathrm h$ are 800~G (red) and 1400~G (yellow). Right: comparison between
the patterns of the smoothed vertical velocity field (grey-scale filled contour map with the dotted contours
for zero velocity) and smoothed vertical magnetic field (colour contours) for the same session;
the contour increment is 800~G for magnetic field and 0.2~km~s$^{-1}$ for velocity; red contours correspond
to $B_\mathrm v >0$, green dot--dashed contours to $B_\mathrm v =0$ and yellow contours to
$B_\mathrm v <0$; the size of the smoothing window is 3.5~Mm for magnetic field and 7.4~Mm for velocity.
The dark areas in both grey-scale maps correspond to negative values (vectors directed upward)
and light areas to positive values (vectors directed downward).}
\label{session1}
\end{figure}

\begin{figure}[h!]
\centering
\includegraphics[width=0.495\textwidth,bb=0 0 377 411,clip]{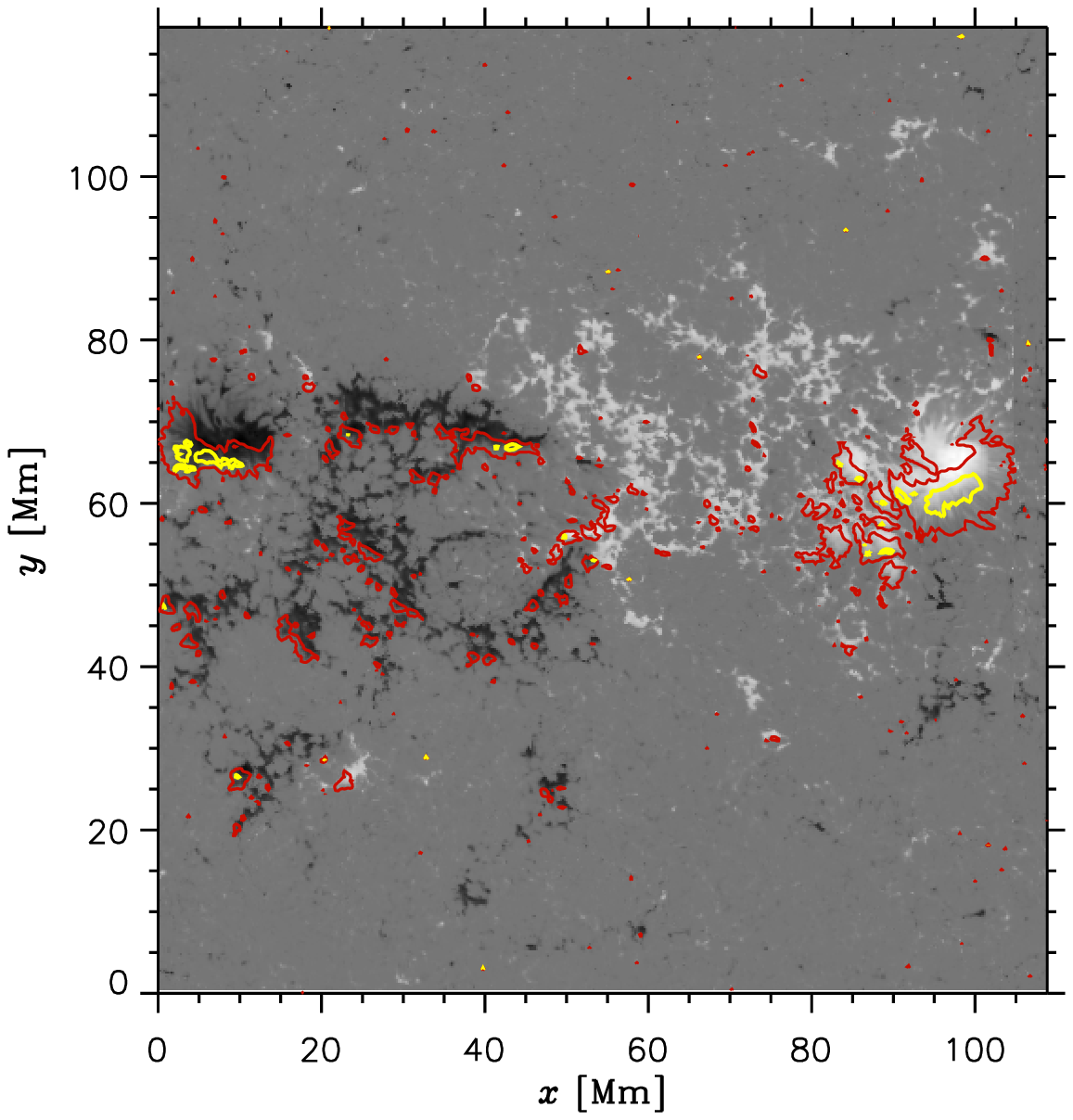}
\includegraphics[width=0.452\textwidth,bb=15 0 360 360,clip]{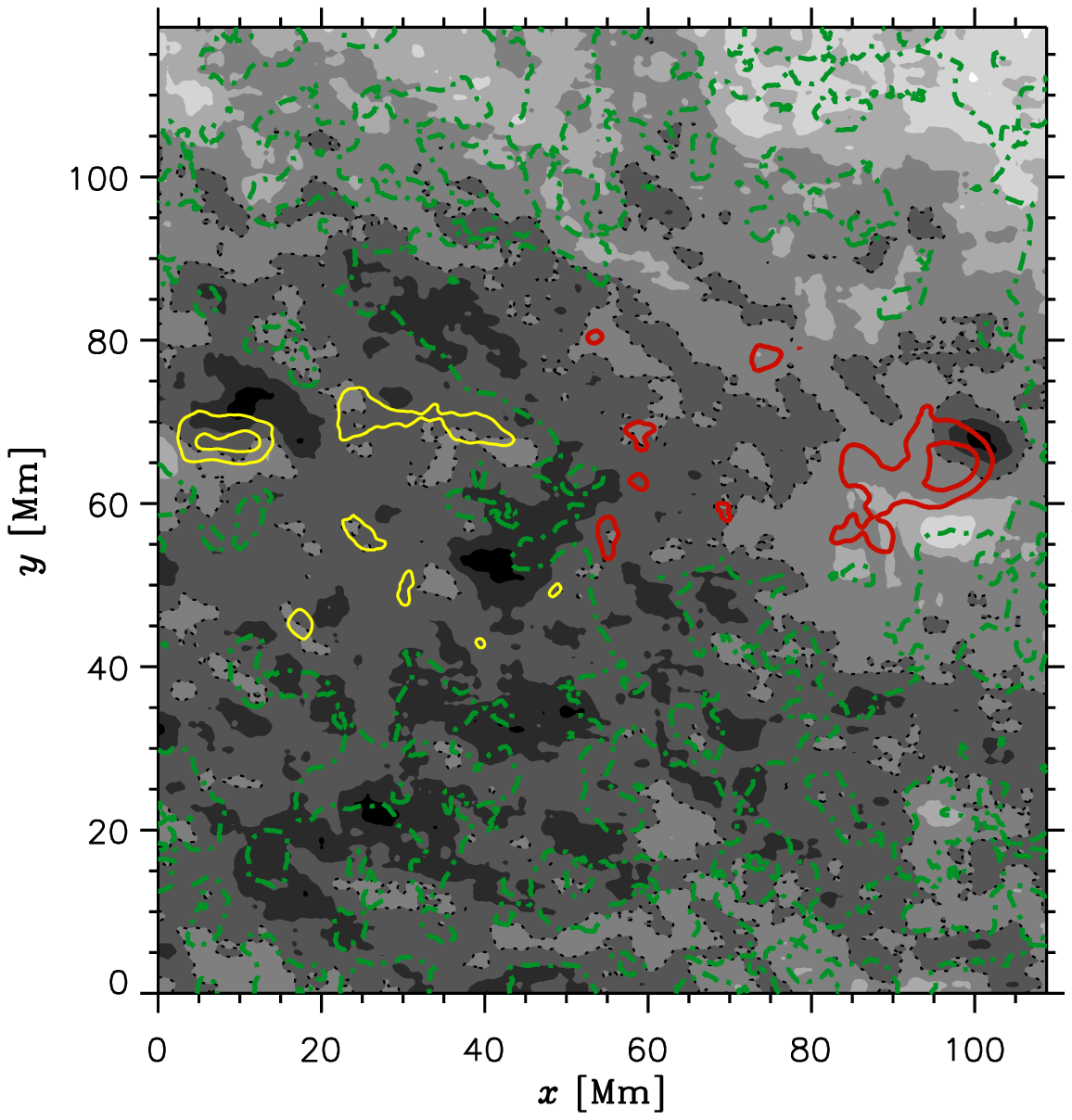}
\caption{Same as in Figure~\ref{session1} but for session (2); the $B_\mathrm v$ range is $[-2121~\mathrm G,\
2460~\mathrm G]$.\protect}
\label{session2}
\end{figure}

\begin{figure}
\centering
\includegraphics[width=0.495\textwidth,bb=0 0 377 411,clip]{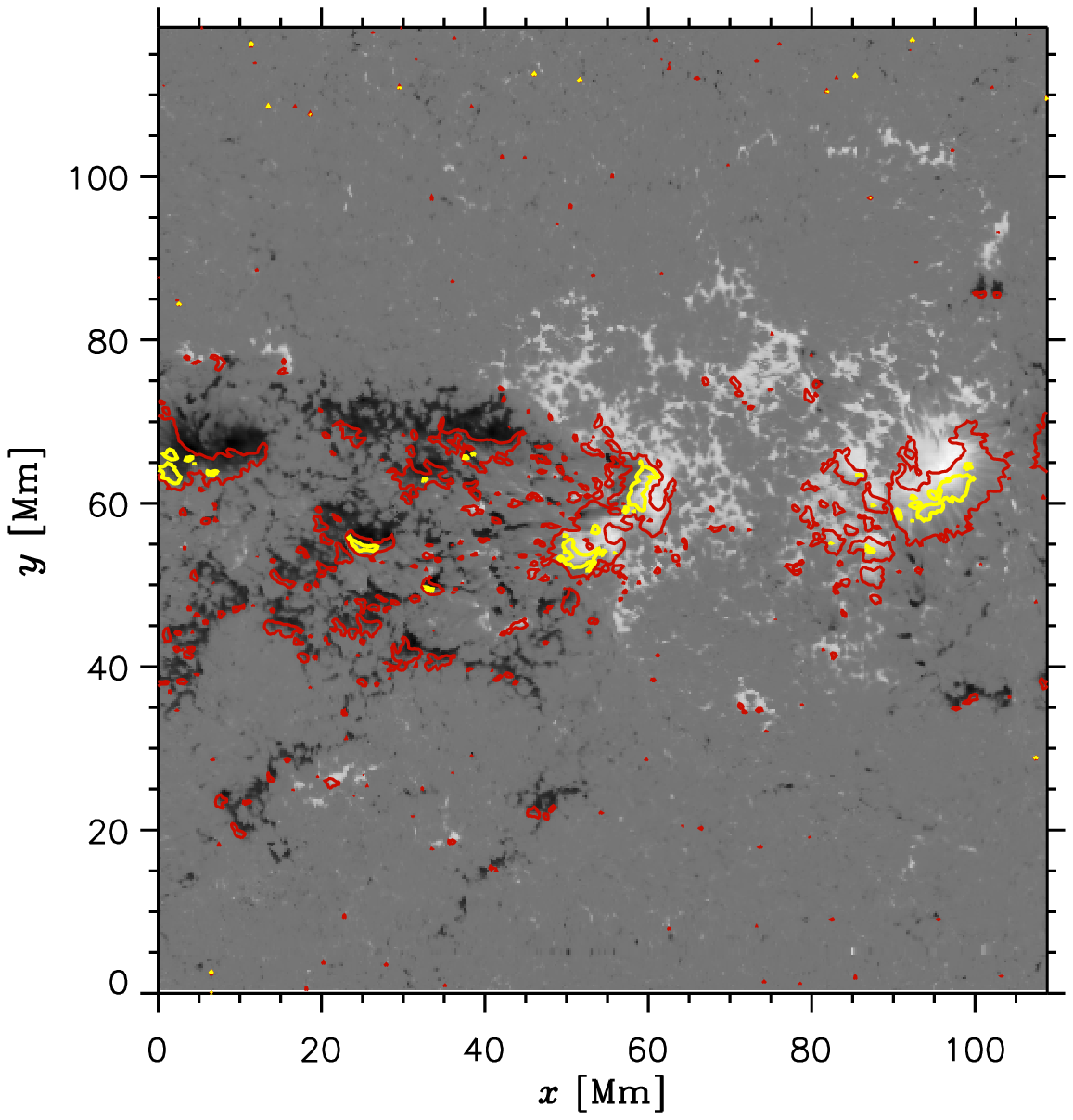}
\includegraphics[width=0.452\textwidth,bb=15 0 360 360,clip]{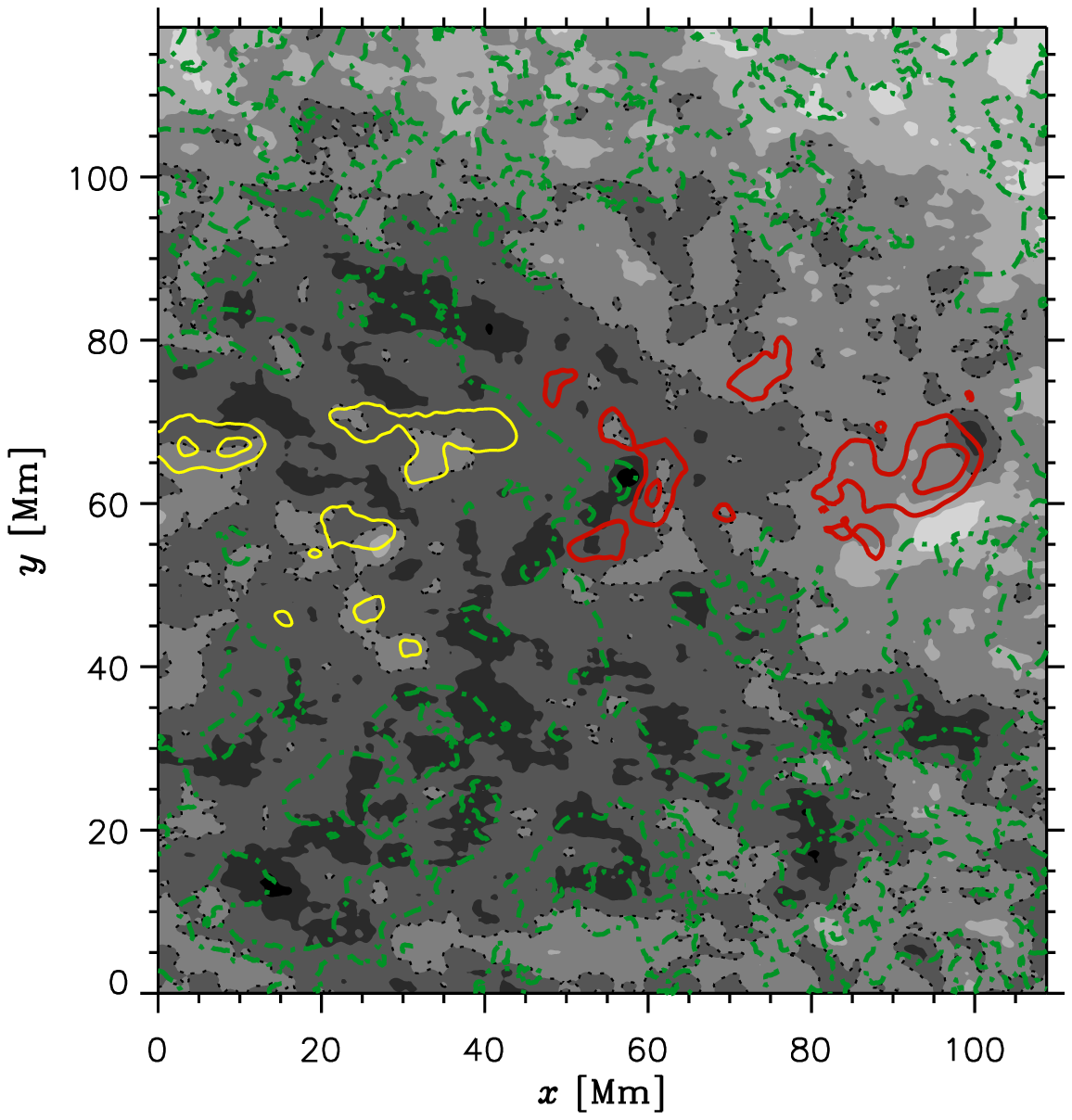}
\caption{Same as in Figure~\ref{session1} but for session (3); the $B_\mathrm v$ range is $[-2112~\mathrm G,\
2467~\mathrm G]$.
\protect}
\label{session3}
\end{figure}

\begin{figure}
\centering
\includegraphics[width=0.495\textwidth,bb=0 0 377 411,clip]{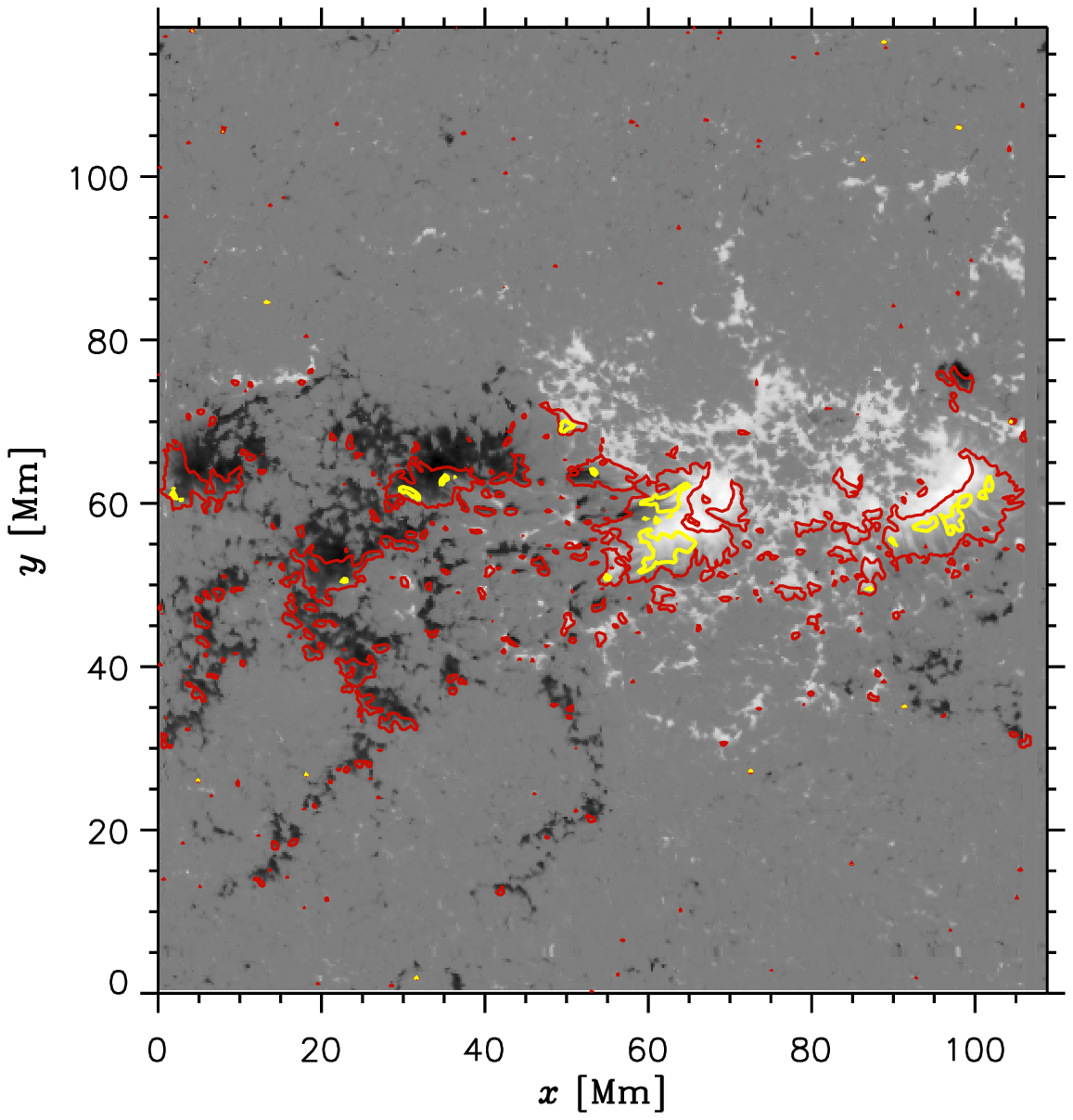}
\includegraphics[width=0.452\textwidth,bb=15 0 360 360,clip]{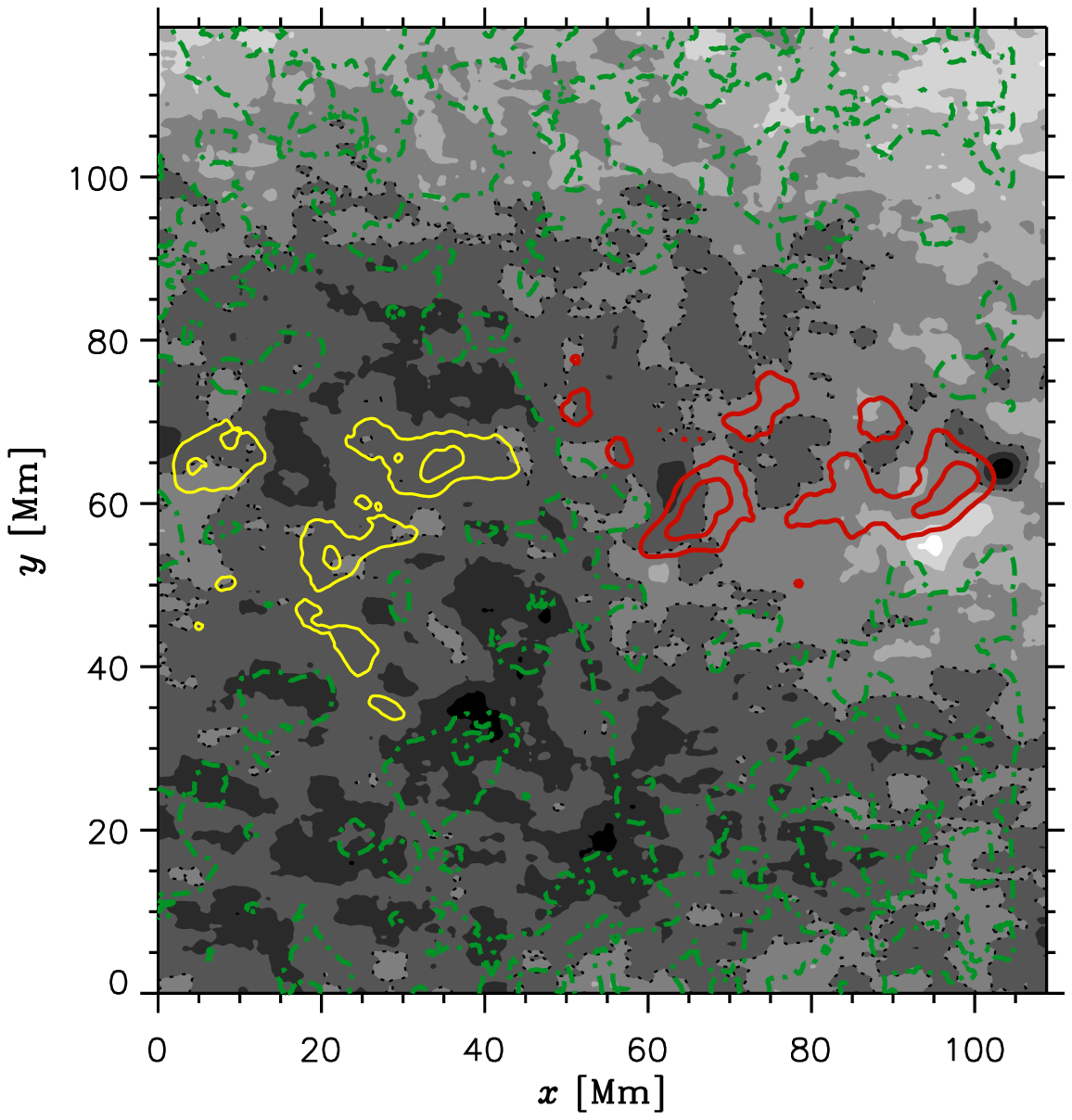}
\caption{Same as in Figure~\ref{session1} but for session (4); the $B_\mathrm v$ range is $[-2177~\mathrm G,\
2217~\mathrm G]$.\protect} \label{session4}
\end{figure}

\begin{figure} 
\centering
\includegraphics[width=0.495\textwidth,bb=0 0 377 411,clip]{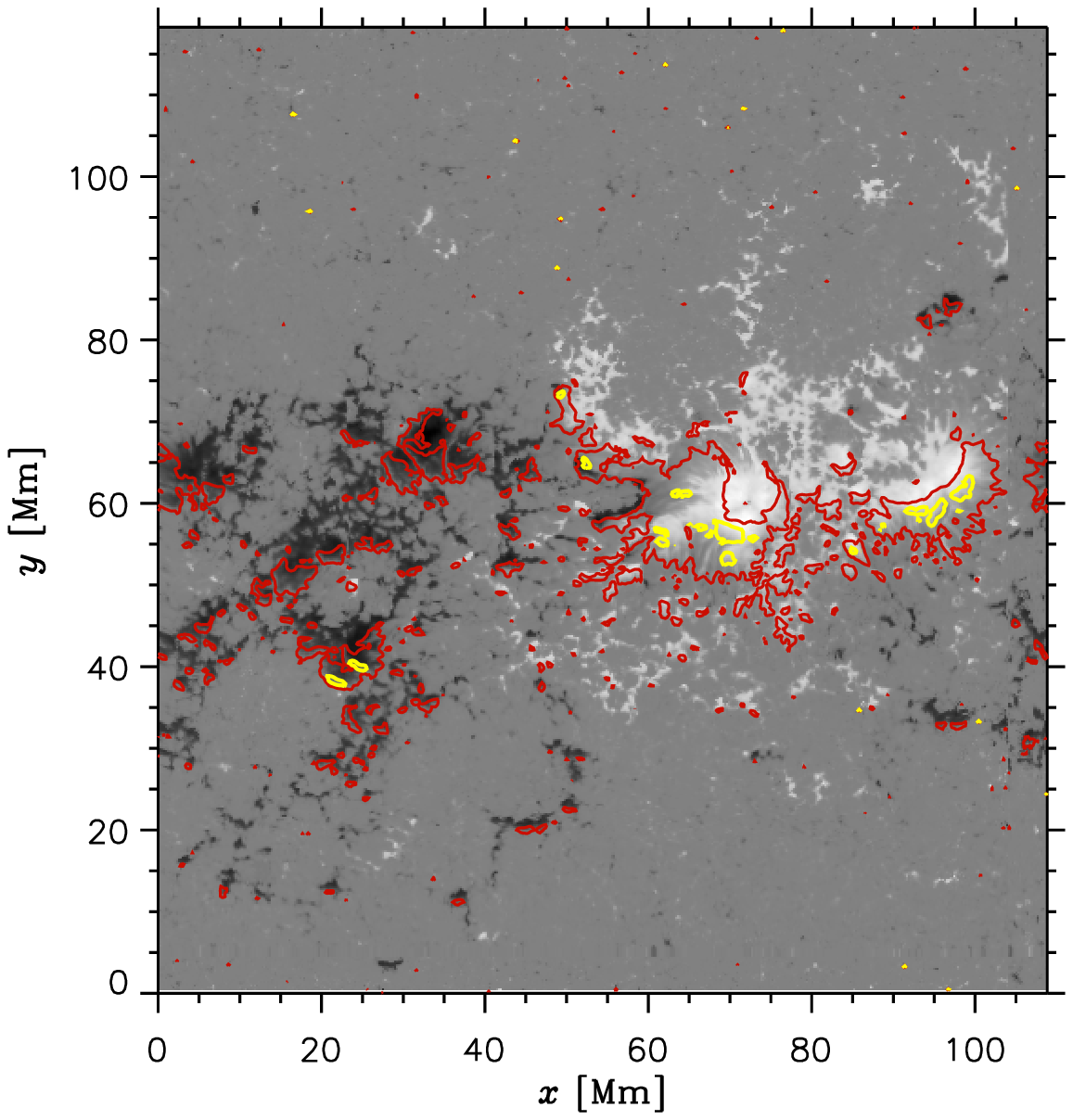}
\includegraphics[width=0.452\textwidth,bb=15 0 360 360,clip]{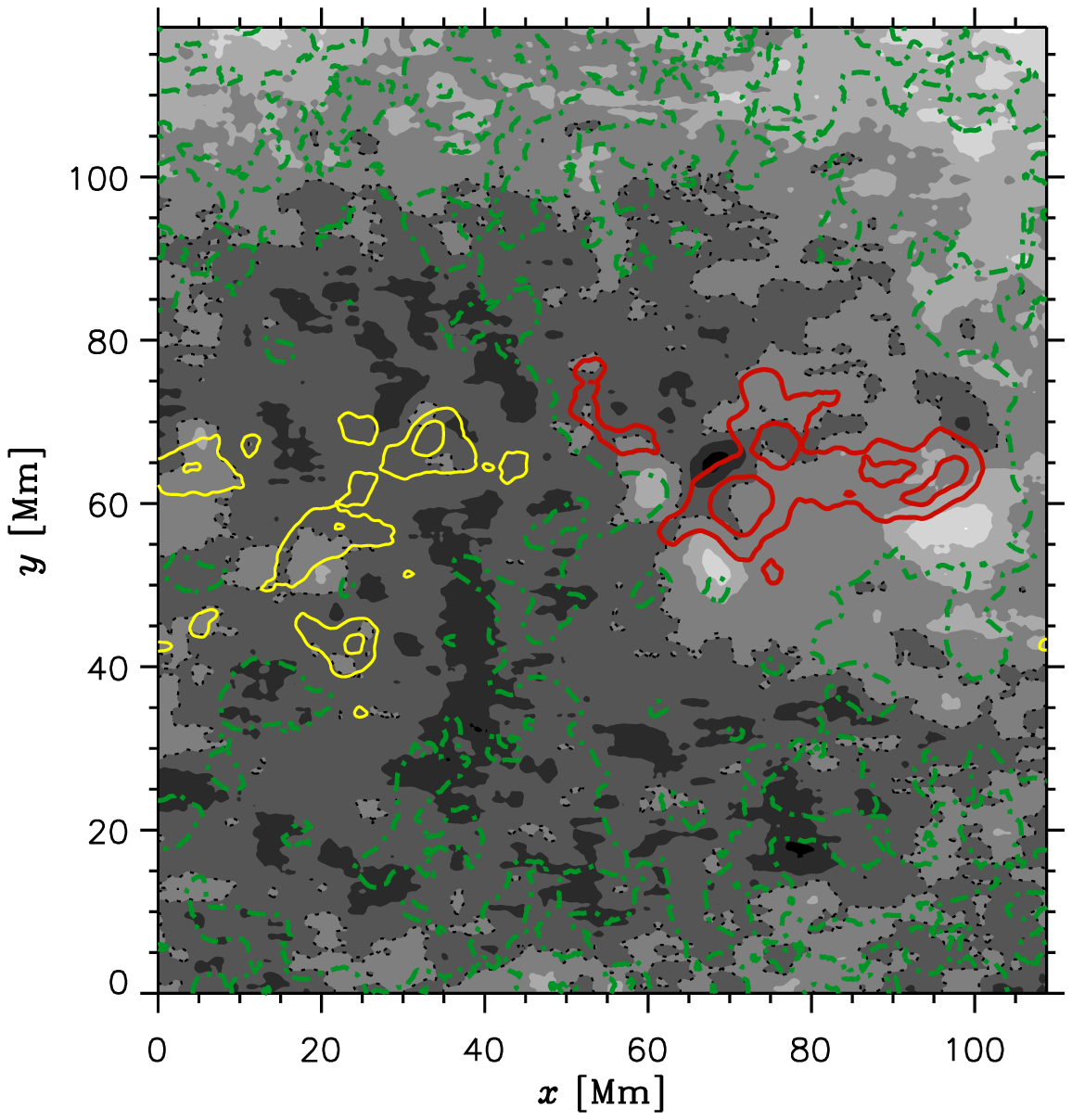}
\caption{Same as in Figure~\ref{session1} but for session (5); the $B_\mathrm v$ range is $[-2362~\mathrm G,\
2316~\mathrm G]$.\protect}
\label{session5}
\end{figure}

\section{Results}

\subsection{Evolution in White Light}

The photospheric images of AR 11313 obtained during the five observational
sessions are shown in Figure~\ref{images}. They cover a time interval somewhat
exceeding one day (see Table~\ref{summary}). It can be seen that a bipolar
sunspot group has already formed by the first observational session. At nearly
the same time, a new, minor sunspot group starts developing between the main
spots of the previously developed group, in the left half of the FOV. This
process becomes mainly accomplished by the third session, after which the spots
of the newly born group continue growing and its leading and trailing spots
move apart.

\subsection{Magnetic-Field Patterns. Bordering Effect}

We consider here some features of the developing magnetic and velocity fields
deducible from Figures~\ref{session1}--\ref{session5}, which refer to sessions
(1)--(5), respectively. The left-hand panel of each figure makes it possible to
compare between the original (detailed) maps of the vertical [$B_\mathrm v$]
and horizontal [$B_\mathrm h$] components of the magnetic field. As for
confronting the vertical magnetic and velocity fields, a specific noise of fine
details in the original maps makes comparisons between them difficult. In view
of this, we use smoothed magnetic and velocity fields for such comparisons
(right-hand panels).

Let us first note a remarkable feature of the magnetic field, which we call the
\emph{bordering effect}. It can be revealed in the magnetograms of all the five
sessions by examining the left-hand panels but is most pronounced in
Figures~\ref{session4} and \ref{session5}. While most local (mainly positive)
extrema of $B_\mathrm v$ spatially coincide with areas of low $B_\mathrm h$,
these areas are partially bordered with arched  areas of locally enhanced
$B_\mathrm h$.

The bordering effect admits a quite straightforward interpretation. Assume that
magnetic field lines form a bundle issuing from the subphotospheric layers and
diverge in its top section like the water jets in a fountain. In this case,
$B_\mathrm v$ will obviously reach its peak magnitudes in the central part of
the bundle, while $B_\mathrm h$ will be maximum in an annulus surrounding the
bundle.

A common feature of all the magnetograms is fairly good, visually noticeable
spatial correlation between the distributions of $B_\mathrm v$ and $B_\mathrm
h$ (see left-hand panels in Figures~\ref{session1}--\ref{session5}). However,
for a given sign of $B_\mathrm v$, its pattern typically exhibits a relatively
uniform spatial shift with respect to the corresponding, similar pattern of
$B_\mathrm h$ (the contour representation of $B_\mathrm h$ clearly visualises
this shift only in the areas of $B_\mathrm v < 0$). The shifts for the two
signs may differ in their magnitude and direction; they do not exceed 2~Mm.

The shift of the $B_\mathrm h$ pattern relative to the $B_\mathrm v$ pattern
may be indicative of a systematic increase in the tilt of the magnetic field
lines in the direction of this shift. This feature thus proves to be akin, in a
sense, to the bordering effect, being a manifestation of the three-dimensional
structure of the magnetic field.

To make further inferences from the magnetograms, we have to remember a feature
that should be expected in the case of the flux-tube rise and was mentioned
above as item (i) in the list of RTM doubtful points (see Introduction).
Specifically, strong horizontal fields should connect the emergence areas of
the main spots, forming elongated features where the horizontal-field strengths
values are comparable with the vertical-field strengths observed in the spots.

As for our magnetograms, they display signatures of finely structured
horizontal field in the form of very narrow elongated features stretching
between the main magnetic poles. These features are present in the obtained
$B_\mathrm h$ distributions for sessions 3, 4 and (in a very faint appearance)
5 but can hardly be distinguished in their contour representations in the
respective figures. In this context, the following should be noted. First,
these features became noticeable after the formation of the bipolar
configuration of strong magnetic field but were completely absent at the stage
of development of this configuration. In contrast, if a flux-tube loop rose, we
would see stretched features (not necessarily finely structured) earlier,
during the sunspot-group development. Second, as the magnetograms demonstrate,
the horizontal field in these features is considerably weaker (a factor of two
to three) than the peak $B_\mathrm h$ values, which, in turn, do not exceed the
peak values of $B_\mathrm v$ (see Table~\ref{values}). Therefore, the scenario
of magnetic-field development is not consistent with the RTM-based expectation
indicated as item (i) in the list of RTM doubtful points (see Introduction).

\begin{table}[h]
\caption{Characteristic values of $B_\mathrm v$ and $B_\mathrm h$ in the
developing AR}\label{values}
\begin{tabular}{cccc}
  \hline
  Session & $B_\mathrm v$ range & $B_\mathrm h$ range & Typical $B_\mathrm h$ values \\
  No.      &  [G]              &  [G]                 &   in fine features [G]\\
  \hline
  1 &    ($-1900$)--2500& 0--1900 & 800--1200
 \\
  2 &   ($-2100)$--2500 & 0--1800 & 600--1000
 \\
  3 &   ($-2100$)--2500 & 0--2100 & 700--1100
 \\
  4 &   ($-2200$)--2200 & 0--1800 & 700--1100 \\
  5 &   ($-2400$)--2300 & 0--2200 & 600--1100
 \\
  \hline
\end{tabular}
\end{table}

\subsection{The Magnetic \textit{vs.} the Velocity Field}

Now let us compare the smoothed $B_\mathrm v$ and $u_\mathrm v$ fields shown in
the right-hand panels of Figures~\ref{session1}--\ref{session5} and discuss the
relationship between them. It can be seen from our velocity maps that no marked
upflow can definitely be related to the entire area where the rise of a
flux-tube loop should be expected according to the RTM. In
Figure~\ref{session1}, several moderate, localised upflows are present in this
area, while the strongest upflows are present not far from the lower left
corner of the FOV and near the existing sunspots. In Figure~\ref{session2}, an
upflow-velocity maximum is located below the area of interest, being comparable
in its magnitude with several other local maxima. Nothing similar to a strong
upflow associated with the hypothetical rising tube can be seen in
Figure~\ref{session3}, not to say about Figures~\ref{session4}
and~\ref{session5}.

It is also worth remembering that the $u_\mathrm h$ field that we constructed
based on the same observations as those discussed here (see Paper~I) exhibits
neither any spreading flow on the scale of the whole developing group nor any
flows qualitatively different from normal mesogranular and supergranular
convection unaffected by any large-scale disturbances.

Thus, the observed patterns of both the vertical and the horizontal velocity in
the developing active subregion are at variance with the RTM-based expectation
mentioned above as item (ii) in the list of RTM doubtful points (see
Introduction).

It is also interesting to note the following particular feature of the
$u_\mathrm v$ field in the vicinity of the leading (rightmost) spot of the
large group that was formed before the smaller group started developing. If we
compare Figure~\ref{images}a--c with $B_\mathrm v$ and $u_\mathrm v$
distributions in the right-hand panels of
Figures~\ref{session1}--\ref{session3}, we can find that, on two sides of the
leading spot, there are an upflow (above the spot in the FOV) and a downflow
area (below it). This feature can also be distinguished (although is much less
pronounced) in Figures~\ref{session4} and \ref{session5}.

\section{Discussion and Conclusion}

Our analysis of the observational data for the development of the minor
subregion within AR 11313 on 9-10 October 2011 reveals a noticeable discrepancy
between the observed development pattern of the magnetic and velocity fields
and the RTM-based expectations.

First, we have found that the distributions of $B_\mathrm v$ and $B_\mathrm h$
over the area of the growing magnetic subregion are spatially well correlated,
with a shift between the entire patterns of $B_\mathrm v$ and $B_\mathrm h$ by
a distance of no larger than about 2~Mm. The rise a flux-tube loop should
result in a qualitatively different pattern. The maxima of the two
magnetic-field components would be spatially separated in this case: the
vertical field would be the strongest where either of the main spots emerges,
with the maximum horizontal-field strengths reached in between them. Moreover,
the horizontal field would form elongated features connecting the main spots,
being comparable with the vertical-field strengths in these spots. In contrast,
we observe finely structured elongated features, which emerged no earlier than
the bipolar configuration of strong magnetic field had completely formed and
were considerably weaker than the vertical field in the spots.

The feature that we call the bordering effect provides additional evidence
against the rise of a tube, since it demonstrates a fountainlike
three-dimensional structure of individual local magnetic-field maxima. It
becomes pronounced by the final formation stage of the active subregion under
study (Figure~\ref{session3}). Such a structure is hardly compatible with the
emergence of a whole tube loop, which should produce local maxima of the
vertical magnetic field only at the feet of the loop.

Second, the flow pattern observed in the area of the developing active
subregion is not consistent with the idea of the flux-tube-loop emergence.
There is no upflow on the scale of the whole subregion, which should be related
to the rising-tube process. In addition, the horizontal-velocity field
(considered in Paper~I) does not exhibit any spreading flow on the scale of the
entire growing magnetic region; instead, some motions resembling normal
mesogranular and supergranular flows appear to be preserved. In this context,
it is worth noting that, many years ago, \citet{bum63,bum} and \citet{bumhow}
found that the growing magnetic fields do not break down the pre-existing
convective-velocity field but come from below ``seeping'' through the network
of convection cells. The development of a sunspot group seems to be controlled
by the supergranular network, and the lines of force of the strong local
magnetic fields are nearly collinear with the streamlines of the photospheric
plasma.  This can naturally be understood if the magnetic field is assumed to
be formed by convective motions.

Thus, our observational data can hardly be interpreted in the framework of the
rising-tube model. However, our inferences are based on a single AR-emergence
event, which is insufficient to assess how typical the observed scenario is.

The emergence of 41 AR is analysed by \cite{Poisson_etal:2015} based on
observations of the vertical magnetic fields. They use a model of the twisted
flux tube (rope) to describe the time variations in the twist (and,
accordingly, in the magnetic helicity). However, as can be observed, the
assumption of the flux-tube rise is the starting point of their investigation
and its results are discussed in comparisons with numerical simulations of the
rising tubes. Since no possibilities lying beyond the framework of this concept
are considered, it is not obvious that such alternatives should be ruled out in
the cases considered.

The issue of the comparative role of different really possible active-region
generation mechanisms thus remains open, and further observations are needed.
In our view, they would be most promising if they include recording both
magnetic and velocity fields. We plan such observations for the near future.
Since the rising-tube mechanism does not appear to be universal, local
convective dynamo as the producer of the magnetic fields of active regions
deserves close attention (see Paper~I for a discussion of the relevance of this
mechanism).

\begin{acks}
\emph{Hinode} is a Japanese mission developed and launched by ISAS/JAXA, with
NAOJ as domestic partner and NASA and STFC (UK) as international partners. It
is operated by these agencies in cooperation with ESA and NSC (Norway). The
work of A.V.G. and A.A.B. was supported by the Russian Foundation for Basic
Research (project no. 12-02-00792-a). We are grateful to L.M. Alekseeva and to
the reviewer for their comments.
\end{acks}

\bibliographystyle{spr-mp-sola}
\bibliography{Getling_etal}

\end{article}
\end{document}